\documentclass[aps,prb,twocolumn,superscriptaddress,showpacs]{revtex4}
\usepackage{graphicx}
\begin{document}

\title{Antiferromagnetic fluctuations in the normal state of LiFeAs}

\author{P. Jegli\v{c}}
\affiliation{Jo\v{z}ef Stefan Institute, Jamova 39, 1000 Ljubljana, Slovenia}

\author{A. Poto\v{c}nik}
\affiliation{Jo\v{z}ef Stefan Institute, Jamova 39, 1000 Ljubljana, Slovenia}

\author{M. Klanj\v{s}ek}
\affiliation{Jo\v{z}ef Stefan Institute, Jamova 39, 1000 Ljubljana, Slovenia}

\author{M. Bobnar}
\affiliation{Jo\v{z}ef Stefan Institute, Jamova 39, 1000 Ljubljana, Slovenia}

\author{M. Jagodi\v{c}}
\affiliation{Institute of Mathematics, Physics and Mechanics, Jadranska 19, 1000 Ljubljana, Slovenia}

\author{K. Koch}
\affiliation{Max-Planck-Institut f\"ur Chemische Physik fester Stoffe, N\"othnitzer Str. 40, 01187 Dresden, Germany}

\author{H. Rosner}
\affiliation{Max-Planck-Institut f\"ur Chemische Physik fester Stoffe, N\"othnitzer Str. 40, 01187 Dresden, Germany}

\author{S. Margadonna}
\affiliation{School of Chemistry, University of Edinburgh, West Mains Road, EH9 3JJ Edinburgh, United Kingdom}

\author{B. Lv}
\affiliation{Department of Chemistry and TCSUH, University of Houston, Houston, TX 77204-5002, USA}

\author{A. M. Guloy}
\affiliation{Department of Chemistry and TCSUH, University of Houston, Houston, TX 77204-5002, USA}

\author{D. Ar\v{c}on}
\affiliation{Jo\v{z}ef Stefan Institute, Jamova 39, 1000 Ljubljana, Slovenia}
\affiliation{Faculty of Mathematics and Physics, University of Ljubljana, Jadranska 19, 1000 Ljubljana, Slovenia}

\date{\today}

\begin{abstract}
We present a detailed study of $^{75}$As NMR Knight shift and spin-lattice relaxation rate in the normal state of stoichiometric polycrystalline LiFeAs. Our analysis of the Korringa relation suggests that LiFeAs exhibits strong antiferromagnetic fluctuations, if transferred hyperfine coupling is a dominant interaction between $^{75}$As nuclei and Fe electronic spins, whereas for an on-site hyperfine coupling scenario, these are weaker, but still present to account for our experimental observations. Density-functional calculations of electric field gradient correctly reproduce the experimental values for both $^{75}$As and $^7$Li sites. 
\end{abstract}

\pacs{74.70.-b, 76.30.-v, 76.60.-k}

\maketitle
%\section{Introduction}

Following the discovery of superconductivity in LaFeAsO$_{1-x}$F$_x$,\cite{Kamihara_2008} nuclear magnetic resonance (NMR) provided one of the earliest evidences for unconventional pairing in the superconducting (SC) state,\cite{Grafe_2008,Nakai_2008,Terasaki_2009} multigap superconductivity,\cite{Kawasaki_2008,Matano_2009,Yashima_2009} pseudogap (PG) behavior in the normal state,\cite{Grafe_2008,Nakai_2008,Imai_2008} and antiferromagnetic (AFM) ordering of Fe$^{2+}$ spins in the undoped parent compounds of Fe-As superconductors.\cite{Nakai_2008,Kitagawa_2008,Jeglic_2009} Although the SC pairing mechanism is still under debate, it is commonly believed that AFM fluctuations play an important role in promotion of high-temperature superconductivity in this family. This is indicated by the presence of the AFM phase next to the SC ground state in the phase diagrams of $RE$FeAsO\cite{Luetkens_2009} ('1111', $RE$ = rare earth) and $A$Fe$_2$As$_2$\cite{Chen_2009} ('122', $A$ = alkaline earth metal) compounds.

Recently, LiFeAs$\--$the so-called '111' member of the Fe-As superconductors$\--$has been reported\cite{Tapp_2008} to undergo a transition to the SC state at $T_c=18$~K without additional doping and apparent AFM ordering or accompanying structural phase transition. Its structure is a simplified analogue of the '1111' or '122' members: FeAs layers comprised of edge-sharing FeAs$_4$ tetrahedra are separated by double layers of Li ions. However, the tetrahedra are deformed and the Fe-Fe distance is considerably shorter compared to other Fe-As superconductors. Moreover, $T_c$ linearly decreases with applied pressure, similarly as in overdoped K$_x$Sr$_{1-x}$Fe$_2$As$_2$, although the charge count of $-1$ per FeAs unit would rather compare LiFeAs to undoped SrFe$_2$As$_2$.\cite{Gooch_2009} LiFeAs is also claimed to be a weakly to moderately,\cite{Kurmaev_2009} or moderately to strongly\cite{Hozoi_2009} correlated system. These conflicting results raise an important question about the significance of AFM fluctuations and the placement of LiFeAs in the general Fe-As superconductor phase diagram.

Here we employ $^{75}$As NMR to quantitatively account for the extent of spin correlations in the normal state of LiFeAs and compare it to a typical '122' member. We find that the spin-lattice relaxation rate $T_1^{-1}$ is enhanced, compared to the values calculated for the noninteracting electron scenario. The quantitative comparison with cuprates and organic superconductors\cite{Slichter_2006} indicates that AFM correlations may also play an important role in the LiFeAs superconductor.

%\section{methods}

\begin{figure}
\includegraphics[width=0.9\linewidth, trim=10 12 5 10, clip=true]{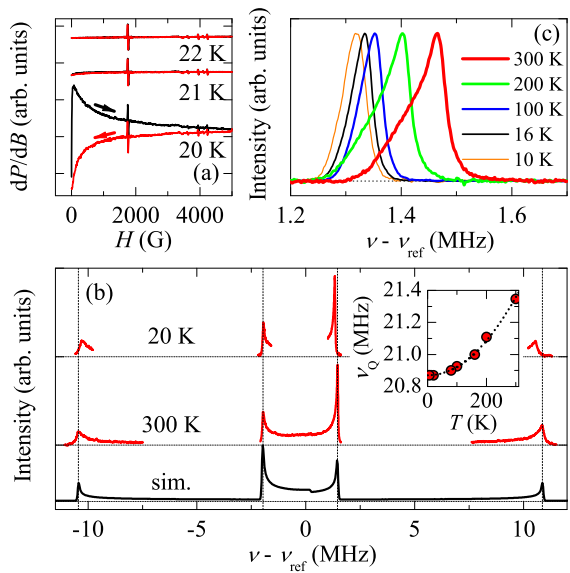}
\caption{(color online). (a) Microwave absorption near the SC transition at low magnetic field indicating $T_c\sim20$~K. Arrows show different field sweep directions. Sharp peaks at around 1700 G originate from a dielectric resonator. (b) $^{75}$As NMR spectrum at $300$~K and $20$~K for a chosen orientation of LiFeAs polycrystalline sample. A comparison with simulated powder spectrum with $\nu_Q=21.35$~MHz and $K_{\rm iso}=0.32\%$ demonstrates that the sample contains at least few tens of grains. Inset shows the experimental $\nu_Q$ as a function of temperature. (c) Temperature dependence of the high-frequency singularity of the $^{75}$As NMR central transition.}
\label{line_As}
\end{figure}

Stoichiometric polycrystalline LiFeAs was synthesized from high-temperature reactions as described in detail in Ref.~\onlinecite{Tapp_2008}. For magnetic resonance experiments the LiFeAs sample was sealed into the quartz tube under vacuum to avoid contamination with moisture during the measurements. To check the quality of our polycrystalline LiFeAs samples, we performed electron paramagnetic resonance (EPR) measurements in the vicinity of SC transition. A non-resonant microwave absorption effect\cite{Miller_1987} occurs sharply below 21 K [Fig.~\ref{line_As}(a)], demonstrating the onset of SC state at $T_c\sim20$~K in agreement with Ref.~\onlinecite{Tapp_2008} and demonstrating the high-quality of our sample.  $^{75}$As ($I=3/2$) NMR frequency-swept spectra were measured in a magnetic field of 9.4~T  with a two-pulse sequence ${\beta}-{\tau}-{\beta}-{\tau}-{\rm echo}$, a pulse length $\tau_{\beta}=5~\rm \mu$s, interpulse delay $\tau = 100~{\rm\mu}$s, and repetition time 100~ms at room temperature. The reference frequency of $\nu (^{75}{\rm As}) = 68.484~$MHz was determined from a NaAsF$_6$ standard. The $^{75}$As $T_1^{-1}$ was measured with inversion-recovery technique. The band structure calculations were performed within the local density approximation (LDA), as described in detail in Refs.~\onlinecite{Jeglic_2009,Grafe_2009}. As basis set Li $(/2s2p3d+3s3p)$, Fe $(3s3p/4s4p3d+5s5p)$ and As $(3s3p3p/4s4p3d+5s5p)$ were chosen for semicore/valence+polarization states. A well converged $k$-mesh with 1183 $k$-points in the irreducible part of the Brillouin zone was used. The structural parameters were taken from Ref.~\onlinecite{Tapp_2008}. The calculated $V_{zz}$ component of the electric field gradient (EFG) tensor is converted into the experimentally measured quadrupole splitting $\nu_Q$ using the relation $\nu_Q=3eV_{zz}Q/[2hI(2I-1)]$, with the quadrupole moment $Q$ and nuclear spin $I$ given in Table~\ref{nuQ}.

%\section{results} 

Representative $^{75}$As NMR spectra of the central $(-\frac {1}{2}\leftrightarrow\frac {1}{2})$ and the satellite $(\pm\frac {3}{2}\leftrightarrow \pm\frac {1}{2})$ transitions for the polycrystalline LiFeAs sample are shown in Fig.~\ref{line_As}(b) for temperatures between room temperature and $T_c$. Over the entire temperature range the line shape remains characteristic for an axially symmetric EFG tensor, in accordance with the $^{75}$As site symmetry $4mm$, indicating the absence of a structural phase transition, as encountered in the undoped '1111' and '122' members of the Fe-As superconductors family. Analysis of the splitting  between both singularities belonging to the satellite transitions reveals only a moderate temperature dependence of $\nu_Q$, which monotonically decreases from 21.35~MHz at room temperature reaching 20.87~MHz at low temperatures [inset to Fig.~\ref{line_As}(c)]. There is no indication of AFM ordering down to $T_c$, which would be seen as an abrupt broadening of the NMR line shape due to the appearance of internal magnetic fields.\cite{Nakai_2008,Kitagawa_2008,Jeglic_2009}

\begin{figure}
\includegraphics[width=0.9\linewidth, trim=10 10 10 10, clip=true]{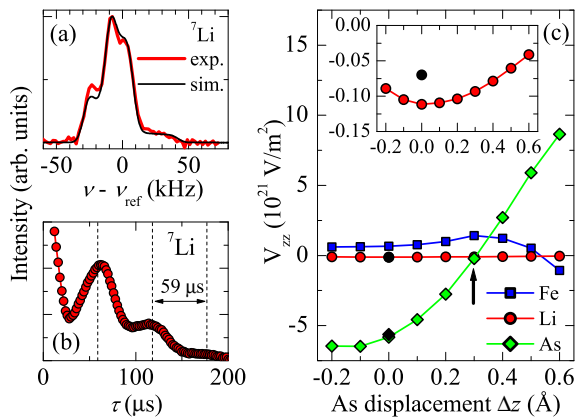}
\caption{(color online). (a) Experimental (thick red line) and calculated (thin black line) $^7$Li NMR spectra at 300~K and magnetic field 4.7~T [$\nu_{\rm ref}(\rm LiCl)=77.7247$~MHz] of LiFeAs polycrystalline sample. (b) $^7$Li echo amplitude as a function of interpulse delay $\tau$ measured at 300~K (see text for details). (c) The calculated $V_{zz}$ at the As (green diamonds), Li (red circles) and Fe (blue squares) sites as a function of $\Delta z$ (see text for details), together with experimental data for Li (black circle) and As (black diamond). The minimum in energy regarding the As $z$ position is marked by the black arrow. The inset shows the Li values on a smaller scale.}
\label{calc_EFG}
\end{figure}

In Fig.~\ref{calc_EFG}(a) we show the $^7$Li ($I=3/2$) NMR spectrum measured at 300 K. Contrary to the $^{75}$As resonance the shift of the $^7$Li NMR line is small and negative [Fig.~\ref{calc_EFG}(a)]. However, the value of $-61(5)$~ppm cannot be attributed to the pure orbital shift (typical values are an order of magnitude smaller), which may indicate an incomplete charge transfer from the Li layer to the FeAs layer. From the $^7$Li NMR linewidth $\delta\nu\approx 90$~kHz, we conclude that $^7$Li has a very small $\nu_Q$. In order to extract $^7$Li $\nu_Q$ we performed an echo-decay measurement. The $^7$Li echo amplitude clearly shows characteristic quadrupole oscillations as a function of interpulse delay $\tau$ in the two-pulse $\beta-\tau-\beta-\tau-{\rm echo}$ experiment [Fig.~\ref{calc_EFG}(b)].\cite{Klanjsek_unpub} Oscillations with the period $t_Q=59$~$\mu$s yield $\nu_Q =2/t_Q\approx 34$~kHz. The $^7$Li ($I=3/2$) NMR lineshape simulation taking into account the quadrupole splitting $\nu_Q=34$~kHz and the magnetic anisotropy of $160(5)$~ppm (both obeying axial symmetry in accordance with the $^{7}$Li site symmetry) fits the experimental NMR spectrum very well [Fig.~\ref{calc_EFG}(a)].

Next we compare the experimental values of quadrupole splittings for $^{75}$As and $^7$Li with those obtained from the band structure calculations. As usually encountered in Fe-As superconductors, the displacement of As site along the $z$ axis has a huge influence on the EFG at the As site, see Fig.~\ref{calc_EFG}(c). Experimental $\nu_Q$ matches the calculated one for $\Delta z=z-z_{\rm exp}=0$, where $z_{\rm exp}=0.2635$ is the experimental As $z$ position\cite{Tapp_2008} (see Table~\ref{nuQ} for details). The minimum in energy with respect to the As $z$ position predicts the displacement of As by almost $\Delta z=0.3$~\AA [marked by the black arrow in Fig.~\ref{calc_EFG}(c)]. The corresponding $\nu_Q\sim 0$ fails to correctly reproduce the measured $^{75}$As $\nu_Q$. This is in line with findings in the '122' compounds \cite{Kasinathan_2009} but in striking contrast to studies of the '1111' compounds, where the calculated and measured $\nu_Q$'s agree well for the optimized As $z$ position.\cite{Jeglic_2009,Grafe_2009} Calculated EFG at the Li site is much smaller, less dependent on the As $z$ position, and does not reach the value $\nu_Q=0$ in the covered interval of $\Delta z$ [inset to Fig.~\ref{calc_EFG}(c)]. This can be understood by different bonding situations: whereas Fe and As build a polyanionic sublattice formed by covalent bonds, Li only has a slightly filled 2$p$ shell. As such, the EFG at the Li site does not provide such a stringent test for the quantity $\Delta z$, in contrast to the EFG at the As site. Anyway, the measured $^7$Li $\nu_Q$ compares relatively well to the range of calculated values.

\begin{table}
\caption{Comparison between calculated and experimental $\nu_{Q}$'s for $^{75}$As and $^7$Li sites. Quadrupole moments $Q$ are taken from Ref.~\onlinecite{Harris_2002}.}
\begin{tabular}{cccccc}
\hline \hline
Site & $I$ & $Q$~(fm$^2$) & $V_{zz}^{\rm calc}$~(V/m$^2$) & $\nu_Q^{\rm calc}$~(MHz) & $|\nu_Q^{\rm exp}|$~(MHz) 
\\ 
\hline 
$^{75}$As & $3/2$& $31.4$ & $-5.82 \cdot 10^{21}$ & $-22.1$ & $21.35$ 
\\ 
$^7$Li & $3/2$ & $-4.01$ & $-0.11 \cdot 10^{21}$ & $0.054$ & $0.034$ 
\\ 
\hline \hline
\end{tabular}
\label{nuQ}
\end{table}

\begin{figure}
\includegraphics[width=0.9\linewidth, trim=10 13 10 10, clip=true]{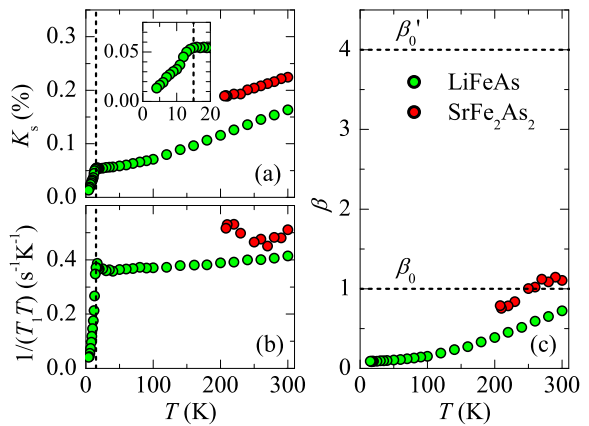}
\caption{(color online). Temperature dependence of the $^{75}$As NMR: (a) spin part of Knight shift, (b) $(T_1T)^{-1}$ and (c) Korringa factor $\beta$ above $T_c$, measured for LiFeAs (green squares) and SrFe$_2$As$_2$ (red circles). Horizontal dashed lines indicate expected values for $\beta$ in case of noninteracting electrons for on-site ($\beta_0$) and transferred coupling ($\beta_0'$) [see text for details]. The inset to (a) shows the behavior of $K_{\rm s}$ below $T_{\rm c}=15$~K (vertical dashed line) at 9.4~T.}
\label{beta}
\end{figure}

We now focus on the role of AFM correlations in LiFeAs. We begin with the determination of the spin part of the $^{75}$As NMR Knight shift from the temperature dependence of the high-frequency singularity of the $^{75}$As central transition [Fig.~\ref{line_As}(c)]. The position of this singularity is given by $\nu = \nu_0(1+K_{\rm iso})+3\nu_Q^2/(16\nu_0)$, where $\nu_0$ is the $^{75}$As Larmor frequency and $K_{\rm iso} = K_{\rm orb}+K_{\rm s}$ represents an isotropic $^{75}$As shift. The latter has two contributions, the orbital part $K_{\rm orb}$ and the spin part $K_{\rm s}$. Taking into account the slight temperature variation of $\nu_Q$ [inset of Fig.~\ref{line_As}(b)] we can extract the precise temperature dependence of $K_{\rm iso}$. For the $^{75}$As orbital contribution we assume $K_{\rm orb}=0.15\%$, which leads to $K_{\rm s}(T \rightarrow 0) = 0$ [inset of Fig.~\ref{beta}(a)] in accordance with the spin-singlet Cooper pairing.\cite{Terasaki_2009} We find that $K_{\rm s}$ is strongly reduced with decreasing temperature and changes from $K_{\rm s}=0.16\%$ to $K_{\rm s}=0.055\%$ between room temperature and $T_{\rm c}=15$~K at 9.4~T [Fig.~\ref{beta}(a)]. Such suppression of $K_{\rm s}$ is reminiscent of the PG behavior observed in many Fe-As superconductors.\cite{Grafe_2008,Nakai_2008,Imai_2008} Because it has been reported for a wide range of $x$ in Ba(Fe$_{1-x}$Co$_x$)$_2$As$_2$,\cite{Ning_2009} the observation of the PG-like behavior is not yet conclusive about the positioning of LiFeAs in the Fe-As superconductor phase diagram.  

We obtain complementary information from the temperature dependence of $^{75}$As spin-lattice relaxation rate $T_1^{-1}$ [Fig.~\ref{beta}(b)]. The nuclear magnetization recovery curves follow $M(t)-M_0\propto 0.1\exp(-t/T_1)+0.9 \exp(-6t/T_1)$ [Ref.~\onlinecite{Matano_2009}] in the whole temperature range. Below 40 K we detect a slight enhancement in $(T_1T)^{-1}$ followed by a sharp decrease below $T_{\rm c}$. However, since $(T_1T)^{-1}$ does not follow the PG-like behavior seen in $K_{\rm s}$, we conclude that AFM fluctuations are present already above 40 K, which is the reason for almost temperature-independent $(T_1T)^{-1}$ above $T_{\rm c}$. Enhancement and divergent behavior of $(T_1T)^{-1}$ due to the slowing down of AFM fluctuations has been reported for underdoped '122' superconductors.\cite{Ning_2009} With increasing doping the AFM fluctuations become less pronounced and $(T_1T)^{-1}$ shows PG behavior in the overdoped regime. Our results suggest that LiFeAs is somewhere in between these two limits with properties analogous to those of optimally doped Fe-As superconductors. It seems that this can explain the relatively high $T_c$, its decrease with the applied pressure and the absence of AFM ordering.    

%\section{discussion}

In order to quantitatively verify the presence of AFM fluctuations in the normal state of LiFeAs, we turn to the analysis of the Korringa relation for $^{75}$As,
\begin{equation}
T_1TK_{\rm s}^2 = \frac{\hbar}{4\pi k_B}\frac{\gamma_e^2}{\gamma_n^2}\beta,
\label{Korringa}
\end{equation}
where $\gamma_{\rm e}$ and $\gamma_{\rm n}$ are the electron and nuclear gyromagnetic ratios, respectively. The phenomenological parameter $\beta$, called the Korringa factor, characterizes the extent of spin correlations.\cite{Pennington_1996_2} In case $^{75}$As couples to the noninteracting Fe $3d$ electrons (i.e., Fermi gas) via the {\it on-site} Fermi contact interaction, the Korringa factor is $\beta=\beta_0=1$. Strong ferromagnetic fluctuations increase the value of $\beta$, while AFM fluctuations decrease it. However, it has been recently proposed for the Fe-As superconductors\cite{Terasaki_2009} that the $^{75}$As nuclei are coupled to the localized Fe electronic spins via the isotropic {\it transferred hyperfine} coupling.\cite{Millis_1990,Mila_1989} According to Millis, Monien and Pines\cite{Millis_1990} this renormalizes the noninteracting $\beta_0$ value. Namely, $T_1^{-1}$ due to the ${\bf q}$-dependent spin fluctuations is obtained from Moriya's expression 
\begin{equation}
\frac{1}{T_1T} \propto \sum_{\bf q}|A({\bf q})|^2 \frac{\chi''({\bf q},\omega_{\rm n})}{\omega_{\rm n}},
\label{Moriya}
\end{equation}
where $\chi''({\bf q},\omega_{\rm n})$ is the imaginary part of the electron spin susceptibility at the wave vector ${\bf q}$ and at the nuclear Larmor frequency $\omega_{\rm n}$. In case $^{75}$As nucleus is coupled to the localized Fe electronic spins via isotropic transferred hyperfine coupling, we have $|A({\bf q})|^2 \propto  \cos^2\frac{q_xa^*}{2} \cos^2\frac{q_ya^*}{2}$, where $a^*$ is the distance between two neighboring Fe$^{2+}$ spins. For  noninteracting spins, $\chi''({\bf q},\omega_{\rm n})$ has no strong singularities in the ${\bf q}$-space, and can be taken out of the summation (integrals) in Eq.~(\ref{Moriya}). Compared to the on-site scenario, we get an extra factor $\int\int {\rm d}q_x {\rm d}q_y / {\int\int {\rm d}q_x {\rm d}q_y \cos^2\frac{q_xa^*}{2} \cos^2\frac{q_ya^*}{2}}=4$, which renormalizes the noninteracting $\beta_0$ value to $\beta_0'=4$. From here we proceed as usual: in case $\beta>4$ ferromagnetic fluctuations are predicted, whereas AFM fluctuations should lead to $\beta<4$. For instance, in cuprates\cite{Millis_1990}$\--$a prototypical example of a system where AFM fluctuations are important$\--$ $\beta$ is reduced by a factor of 15, compared to the noninteracting electron scenario with transferred hyperfine coupling. A similar factor is found in some organic superconductors.\cite{Slichter_2006}

The experimentally extracted Korringa factor $\beta$ for $^{75}$As in LiFeAs is displayed in Fig.~\ref{beta}(c). It amounts to $\sim 0.7$ at room temperature, and then monotonically reduces to $\sim 0.1$ approaching $T_c$. We stress that the absolute values of $\beta$ depend on our choice of $K_{orb}$. For $K_{orb}=0.13\%$ and $K_{orb}=0.17\%$ the low-temperature value of $\beta$ changes to $0.17$ and $0.03$, respectively. Regardless of this uncertainty, the analysis above demonstrates the enhancement of $T_1^{-1}$  at low temperatures with respect to noninteracting electron limits in {\it both} scenaria considered above, and demonstrates the strength of AFM fluctuations in LiFeAs. For comparison we add $\beta$ values for SrFe$_2$As$_2$\cite{Kitagawa_2009} to Fig.~\ref{beta}. In this case, $\beta$ is systematically larger by a factor of $\sim 1.6$ compared to LiFeAs, and above $250$~K $\beta$ is larger than $\beta_0$. In case of the hyperfine transferred coupling scenario, the experimental $\beta$ should be compared to $\beta_0'$ rather than to $\beta_0$. Then, the enhancement of $T_1^{-1}$ in LiFeAs for a factor as large as $40 \pm 20$ at low temperatures suggests strong AFM fluctuations, as recently predicted by quantum chemical calculations.\cite{Hozoi_2009} However, our LDA calculations, which correctly predict $\nu_Q$ for both $^{75}$As and $^7$Li sites without taking into account strong electronic correlations, speak against well defined localized moments at the Fe sites as assumed in the transferred hyperfine coupling scenario. In this case, the correct reference valid for the on-site coupling is $\beta_0=1$ and the enhancement of $T_1^{-1}$ in LiFeAs is reduced to a factor of $10 \pm 5$ speaking for weaker AFM fluctuations. It is not clear at the moment how strongly $\beta$ is enhanced since cross-terms between different bands in the LiFeAs multiband structure can influence $T_1^{-1}$ values,\cite{Walstedt_1994} while they do not affect NMR Knight shifts, so that we cannot unambiguously discriminate between the on-site Fermi contact and the transferred coupling mechanisms. The ambiguity in the analysis above opens three important issues, which will have to be addressed in future studies: (i) Is the coupling of $^{75}$As to itinerant electrons in LiFeAs really on-site, while it is transferred in '122' members? (ii) If this is the case, is it related to structural differences of the FeAs layer between the two families? And, (iii) should LiFeAs really be treated as a strongly correlated system?

%\section{summary}

In summary, NMR and band structure investigations were employed to investigate the normal state properties of the LiFeAs superconductor. The presence of a PG in the uniform spin susceptibility measured by the $^{75}$As Knight shift is overshadowed by AFM fluctuations in the $T_1^{-1}$ measurements.  Although the precise determination of the strength of AFM fluctuations should be a subject of further investigations, we believe that LiFeAs is the simplest Fe-As superconductor where correlation effects might be important and should be considered in future studies.

%\begin{acknowledgments}
We acknowledge stimulating discussions with P. Prelov\v{s}ek, I. Sega and D. Mihailovi\'{c}. This work was supported in part by the Slovenian Research Agency. A. M. G. and B. L. acknowledge the NSF (CHE-0616805) and the R. A. Welch Foundation (E-1297) for support.
%\end{acknowledgments}

\end{document}